\documentclass[acmsmall,screen,nonacm]{acmart}

\usepackage{booktabs}
\usepackage{graphicx}
\graphicspath{{fig/}}

\usepackage{placeins}
\usepackage{tikz}
\usetikzlibrary{arrows.meta,positioning}
\usepackage{xcolor}
\usepackage{framed}
\usepackage{needspace}
\usepackage{enumitem}
\usepackage{wrapfig}
\definecolor{scqTeal}{HTML}{1B6B6B}

\usepackage{listings}
\usepackage{newfloat}
\DeclareFloatingEnvironment[fileext=lol,name=Listing,placement=tbp]{listing}
\lstdefinestyle{scq}{%
  language=Python,
  basicstyle=\ttfamily\scriptsize,
  identifierstyle=\ttfamily\scriptsize,
  keywordstyle=\ttfamily\scriptsize,
  commentstyle=\ttfamily\scriptsize,
  stringstyle=\ttfamily\scriptsize,
  showstringspaces=false,
  columns=fullflexible,
  keepspaces=true,
  numbers=left,
  numberblanklines=false,
  numberstyle=\scriptsize\color{black!55},
  numbersep=7pt,
  xleftmargin=15pt,
  aboveskip=0pt,
  belowskip=0pt,
  frame=none,
}
\newcommand{\lsttop}{\vspace{1pt}\hrule height0.7pt\vspace{3pt}}
\newcommand{\lstbot}{\vspace{3pt}\hrule height0.7pt}

\newcommand{\step}[1]{\textbf{#1}}

\definecolor{shadecolor}{gray}{0.93}
\newcommand{\rqanswer}[2]{%
\par\medskip
\Needspace*{5\baselineskip}%
\begin{snugshade}%
\noindent\textbf{Answer to #1.} #2%
\end{snugshade}%
\par\medskip}

\setcopyright{none}
\renewcommand\footnotetextcopyrightpermission[1]{}

\begin{document}


\title{When Equivalent Quantum Circuits Lose Synthesis Choices}

\author{Boshuai Ye}
\affiliation{%
  \institution{Software Engineering and Information Systems (SEIS),
  Faculty of Information Technology and Electrical Engineering,
  University of Oulu}
  \city{Oulu}
  \country{Finland}
}
\email{boshuai.ye@oulu.fi}

\author{Peng Liang}
\affiliation{%
  \institution{School of Computer Science, Wuhan University}
  \city{Wuhan}
  \country{China}
}
\email{liangp@whu.edu.cn}

\author{Arif Ali Khan}
\affiliation{%
  \institution{Software Engineering and Information Systems (SEIS),
  Faculty of Information Technology and Electrical Engineering,
  University of Oulu}
  \city{Oulu}
  \country{Finland}
}
\email{arif.khan@oulu.fi}

\renewcommand{\shortauthors}{Ye, Liang, and Khan}


\begin{abstract}
Quantum compilers synthesize high-level operations, such as the quantum Fourier transform or multi-controlled X gates, into gate-level circuits, often choosing among synthesis methods that produce different gate counts. Between tools and compilation stages, a circuit may be serialized, exchanged as OpenQASM text, converted between compilers, or lowered to gates. Such a representation change can preserve the computation but leave only a fixed gate-level implementation of the high-level operation. The receiving compiler then loses \emph{synthesis availability}, the ability to apply synthesis to that operation, and compiles the fixed implementation instead. This loss is easy to miss: compilation still succeeds, and functional-equivalence checks still pass.
 
We study synthesis availability in Qiskit, TKET, and Cirq using controlled experiments, repository analysis, circuits captured from project tests, and a Munich Quantum Toolkit (MQT) pipeline. We find that synthesis availability survives serialization when deserialization restores the high-level operation, survives compiler conversion only when the receiving compiler supports that operation, and does not survive the evaluated OpenQASM routes. After a direct OpenQASM~3 round trip, requested synthesis has no effect in all 360 delivered conditions, yet compilation succeeds. The loss changes the compiled circuit's two-qubit gate count, up or down, only where the lost synthesis differs from the fixed implementation; in the MQT pipeline, which requests no synthesis, OpenQASM~2 interchange raises this count by 37.2\% for an eight-qubit Grover circuit.
 
To restore synthesis availability, we introduce \emph{Verified}, which records each high-level operation before the representation change and, afterwards, reconstructs the operation only if a check confirms that the record still matches the delivered circuit. Verified restores the requested synthesis in all 830 recovery runs, with a median verification time of 2.0--638~ms per operation family. The check rejects every record that later transformations have made stale, but also some valid records; reconstructing from stale records without the check produces non-equivalent circuits. Preserving the computation is therefore not enough: compilers and interchange tools should report whether requested synthesis takes effect and verify operation records before reconstruction.
\end{abstract}

\maketitle

\section{Introduction}\label{sec:intro}

Before a quantum program can run on specific hardware, a compiler translates it into supported operations, selects gate implementations, maps logical qubits to physical qubits, and inserts routing operations required by the hardware connectivity~\cite{qiskit2024,tket2020}. These decisions can substantially affect the resulting circuit's gate count and depth~\cite{li2019sabre}.

Quantum software frameworks such as Qiskit, TKET, and Cirq provide high-level operations for expressing complex computations~\cite{qiskit2024,tket2020,cirq}. Examples include the quantum Fourier transform (QFT), multi-controlled X (MCX), and linear reversible functions. A high-level operation specifies a computation without fixing its gate-level realization~\cite{qiskithls,tket2020}. \emph{Synthesis} replaces such an operation with a gate-level implementation. For example, Qiskit provides multiple synthesis methods for QFT and MCX, and the chosen method can depend on hardware connectivity, available spare qubits (\emph{ancillas}), and the desired circuit cost~\cite{kms2007,khattar2024,pmh2008,qiskithls,tket2020}. Prior work has shown that retaining high-level operations can enable smaller compiled circuits~\cite{rosa2025decomp,rosa2025hli}. What remains unclear is what happens when a circuit is saved, exchanged, or converted in a way that preserves the computation but removes the high-level operation before later compilation.

During quantum program development and compilation, a circuit may be serialized and restored using QPY (Qiskit's binary circuit serialization format), exchanged as OpenQASM~3 text and reparsed by the receiving compiler, converted between compilers such as Qiskit and TKET, or lowered to gate-level operations~\cite{openqasm3paper,pytketqiskit,qiskit2024,qiskitqpy,tket2020}. These representation changes support circuit exchange, interoperability between compiler frameworks, and multi-stage compilation, but they can also remove high-level operations that enable the receiving compiler to choose more efficient implementations. We call the point where a receiving compiler obtains the resulting representation a \emph{representation boundary}. 

For example, a direct OpenQASM~3 round trip exports a Qiskit circuit containing a high-level \texttt{QFTGate} to OpenQASM~3 text and immediately imports it back. The imported circuit still computes the QFT, but the \texttt{QFTGate} has become a generic \texttt{Gate} named \texttt{qft} (Figure~\ref{fig:mechanism}A). Compilation still succeeds, yet the requested synthesis method \texttt{qft.line} no longer takes effect, and the median two-qubit gate count rises from 135 to the default 150 (see Section~\ref{sec:motivation}). We call this a loss of \emph{synthesis availability}: the receiving compiler can no longer apply synthesis to the high-level operation.

This loss is easy to miss. Qiskit reports neither errors nor warnings, and a functional-equivalence check tests whether the computation is preserved~\cite{qcec2021,peham2022zx} but does not test whether synthesis remains available. Yet the loss can add two-qubit gates, whose error rates on noisy hardware can be an order of magnitude higher than those of single-qubit gates~\cite{murali2019noise}, reducing the likelihood of a correct result. The loss also occurs without an explicit synthesis request: in a Munich Quantum Toolkit (MQT) pipeline, an OpenQASM~2 round trip removes MCX synthesis availability for an eight-qubit Grover circuit; its two-qubit gate count rises by 37.2\% and its estimated fault-tolerant runtime by 68.0\%. This motivates our first research question (RQ1): which representation changes preserve synthesis availability, and when does losing it affect compiled circuit cost?

Keeping the high-level operation would avoid the loss, but an OpenQASM round trip keeps only a gate's name and definition~\cite{qiskitqasm3import}, and TKET has no comparable QFT operation~\cite{pytketdocs}. The operation must then be reconstructed after the boundary, as some project code already does (see Section~\ref{sec:rq2}). An \emph{operation record} stored with the circuit can keep the operation's type, parameters, and qubit binding. Later transformations such as qubit routing and optimization can, however, leave the record out of date (\emph{stale}), and reconstructing from a stale record can produce a non-equivalent circuit, as it does for all 17 stale records on which we test unchecked reconstruction (see Section~\ref{sec:rq2:stale}). Recovery must therefore check that the record still matches the delivered circuit and otherwise leave the circuit unchanged (Figure~\ref{fig:mechanism}B). This motivates our second research question (RQ2): can a retained operation record restore synthesis availability lost at a representation boundary while preserving functional equivalence, and how much time does verification require?

\definecolor{scqBrick}{HTML}{8E3B2F}
 
\begin{figure}[t]
\centering
\begin{tikzpicture}[
  font=\footnotesize, >={Latex[length=1.6mm,width=1.2mm]},
  every node/.append style={execute at begin node={%
      \hyphenpenalty=10000 \exhyphenpenalty=10000 \hbadness=10000}},
  res/.style={anchor=west, text width=34mm, align=left, font=\scriptsize},
  box/.style={draw=black!60, line width=0.5pt, fill=white,
              inner xsep=3pt, inner ysep=2.5pt, align=left, font=\scriptsize},
  obj/.style={box, text width=18mm, align=center, font=\scriptsize\ttfamily},
  head/.style={font=\footnotesize\bfseries, text=black, anchor=west},
  side/.style={font=\scriptsize, text=black!60, inner sep=1pt},
  ar/.style={->, draw=black!65, line width=0.6pt},
  ln/.style={draw=black!65, line width=0.6pt},
  dot/.style={circle, fill=black!65, inner sep=0pt, minimum size=0.85mm}
]
 
\node[head] at (-17mm,14.2mm) {A\, Loss of synthesis availability};
 
\node[side, anchor=south] at (-5.9mm,9.4mm) {high-level operation};
\node[obj] (typed) at (-5.9mm,6.0mm) {QFTGate};
 
\node[side, anchor=south] at (37mm,9.4mm) {generic operation};
\node[obj] (gen) at (37mm,6.0mm) {Gate("qft")};
 
\draw[ar] (typed.east) --
  node[side, above, text=black, inner sep=2.4pt]
  {direct OpenQASM~3}
  node[side, below, text=black, inner sep=2.4pt]
  {round trip}
  (gen.west);
 
\node[side, anchor=north, text=scqTeal] at ([yshift=-0.6mm]typed.south)
  {synthesis available};
\node[side, anchor=north, text=scqBrick] at ([yshift=-0.6mm]gen.south)
  {synthesis unavailable};
 
\coordinate (fa) at (52mm,6.0mm);
 
\node[res, text=scqTeal] (ok) at (62mm,10.0mm)
  {QFT computation preserved};
 
\node[res, text=scqBrick] (bad) at (62mm,1.8mm)
  {\texttt{qft.line} returns no circuit\\
   no error or warning\\
   two-qubit gates: 135 $\rightarrow$ 150};
 
\draw[ln] (gen.east) -- (fa);
\draw[ar] (fa) |- ([xshift=-4.8mm]ok.west);
\draw[ar, draw=scqBrick] (fa) |- ([xshift=-4.8mm]bad.west);
\node[dot] at (fa) {};
 
\draw[scqTeal, line width=0.7pt]
  ([xshift=-3.4mm,yshift=-0.2mm]ok.west)
  -- ++(0.6mm,-0.7mm) -- ++(1.2mm,1.5mm);
 
\draw[scqBrick, line width=0.7pt]
  ([xshift=-3.5mm,yshift=-0.7mm]bad.west)
  -- ++(1.6mm,1.4mm);
 
\draw[scqBrick, line width=0.7pt]
  ([xshift=-1.9mm,yshift=-0.7mm]bad.west)
  -- ++(-1.6mm,1.4mm);

\node[head] at (-17mm,-4.0mm) {B\, Verified recovery};
 
\node[side, anchor=south west] at (-16mm,-8.6mm)
  {operation record};
 
\node[box, anchor=north west] (rec) at (-16mm,-9.0mm)
  {QFT + parameters\\
   qubits\\
   region fingerprint};
 
\node[box, anchor=west] (del) at (-16mm,-24.2mm)
  {delivered circuit};
 
\node[box, text=scqTeal, align=center, anchor=center] (ver)
  at (33mm,-17.5mm)
  {verify operation record\\
   against delivered circuit};
 
\coordinate (spine) at ([xshift=-7mm]ver.west);
 
\draw[ln] (rec.east) -| (spine);
\draw[ln] (del.east) -| (spine);
\draw[ar] (spine) -- (ver.west);
\node[dot] at (spine) {};
 
\coordinate (fb) at (52mm,-17.5mm);
 
\node[res, text=scqTeal] (good) at (62mm,-11.8mm)
  {verification succeeds\\
   reconstruct \texttt{QFTGate}\\
   requested synthesis restored};
 
\node[res, text=black!55] (skip) at (62mm,-24.0mm)
  {verification fails\\
   keep delivered circuit};
 
\draw[ln] (ver.east) -- (fb);
\draw[ar, draw=scqTeal] (fb) |- ([xshift=-4.8mm]good.west);
\draw[ar, draw=black!45] (fb) |- ([xshift=-4.8mm]skip.west);
\node[dot] at (fb) {};
 
\draw[scqTeal, line width=0.7pt]
  ([xshift=-3.4mm,yshift=-0.2mm]good.west)
  -- ++(0.6mm,-0.7mm) -- ++(1.2mm,1.5mm);
 
\draw[black!55, line width=0.7pt]
  ([xshift=-3.5mm,yshift=-0.7mm]skip.west)
  -- ++(1.6mm,1.4mm);
 
\draw[black!55, line width=0.7pt]
  ([xshift=-1.9mm,yshift=-0.7mm]skip.west)
  -- ++(-1.6mm,1.4mm);
 
\end{tikzpicture}
 
\caption{The QFT example. (A) A direct OpenQASM~3 round trip preserves the computation but removes synthesis availability. (B) Verified reconstructs the operation only if the record matches the delivered circuit.}
\label{fig:mechanism}
\end{figure}
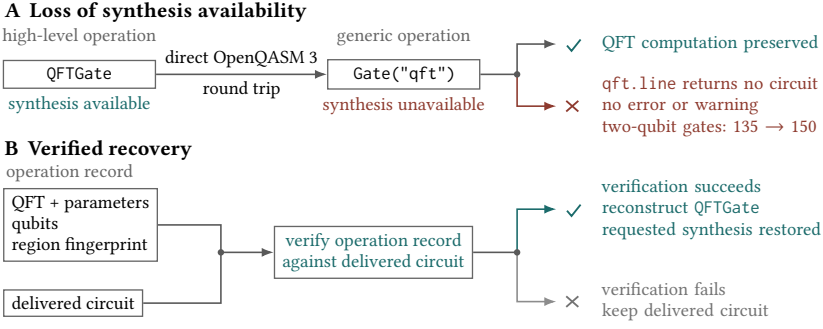


We study both questions in Qiskit, TKET, and Cirq with the following \textbf{contributions}:
\begin{itemize}

\item We distinguish \emph{synthesis availability} from functional equivalence and characterize its preservation across serialization, OpenQASM interchange, compiler conversion, and lowering in three quantum compiler frameworks.


\item We audit repositories of Qiskit-based projects. Their code changes a circuit's representation on 51 code paths, selects a non-default synthesis method in 64 places, and reconstructs a high-level operation in five places. In circuits captured from the projects' tests, OpenQASM interchange raises the two-qubit gate count of 38 MCX circuits by 10--164\%; all 38 circuits have a clean ancilla (a spare qubit in a known state) available at the MCX. OpenQASM interchange also raises the count of five small adders by 25--57\% without any ancilla.

\item We introduce and evaluate \emph{Verified}, which checks an operation record before reconstructing a high-level operation. Verified restores the requested synthesis in all 830 controlled recovery runs and rejects every stale record in our evaluation. 


\end{itemize}

\section{A Motivating Example}\label{sec:motivation}

Consider a developer compiling a ten-qubit quantum Fourier transform (QFT) in Qiskit for a device with linear nearest-neighbor connectivity. Qiskit's default QFT synthesis, \texttt{qft.full}, produces a median of 150 two-qubit gates across ten seeds of the randomized routing pass (\emph{routing seeds}). The developer instead selects \texttt{qft.line}, which is designed for linear connectivity~\cite{qiskithls} and produces a median of 135. This explicit choice reduces the median two-qubit gate count by 10.0\%.

After a direct OpenQASM~3 round trip, compilation still succeeds and the QFT computation is preserved, but the median two-qubit gate count returns to 150, which removes the reduction achieved by \texttt{qft.line}. Table~\ref{tab:example} compares three \emph{routes}, each a different way of changing the circuit's representation.

\FloatBarrier
\Needspace*{10\baselineskip}%
\addvspace{7pt}
\begingroup
\captionof{table}{What each route produces for the ten-qubit QFT on the linear target, using ten-seed medians}
\label{tab:example}
\small
\begin{tabular}{@{}lllr@{}}
\toprule
Route & Representation received & \texttt{qft.line} outcome & Two-qubit gates \\
\midrule
\step{R1} \ QPY round trip
  & \texttt{QFTGate}
  & circuit produced
  & 135 \\

\step{R2} \ Direct OpenQASM~3
  & generic \texttt{Gate} named \texttt{qft}
  & invoked; no circuit
  & 150 \\

\step{R3} \ Explicit lowering
  & elementary gates, no QFT operation
  & not invoked
  & 215 \\
\bottomrule
\end{tabular}
\par\addvspace{7pt}
\endgroup

Neither successful compilation nor functional equivalence reveals that the requested synthesis no longer takes effect. To attribute this loss to the representation change, we keep the QFT computation, target connectivity, compilation settings, synthesis request, and routing seeds fixed and vary only how the representation changes before recompilation.

\emph{\step{R1} QPY round trip.}
QPY restores a \texttt{QFTGate}, so \texttt{qft.line} retains the 135-gate result~\cite{qiskitqpy}.

\emph{\step{R2} Direct OpenQASM~3.}
After re-import, Qiskit represents the imported gate definition as a generic \texttt{Gate} named \texttt{qft}~\cite{qiskitqasm3import}. Qiskit uses the operation name to select QFT synthesis methods, so \texttt{qft.line} is still invoked. The method, however, requires a \texttt{QFTGate}, because user-defined gates may also be named \texttt{qft}~\cite{qiskithls}. For the generic \texttt{Gate}, \texttt{qft.line} returns no synthesis circuit, and compilation continues with the imported gate definition, producing 150 two-qubit gates (Figure~\ref{fig:mechanism}A).

\emph{\step{R3} Explicit lowering.} Decomposition removes the QFT operation, so Qiskit can no longer apply QFT synthesis; the compiled circuit contains a median of 215 two-qubit gates. The loss is harder to detect in R2 because Qiskit still invokes \texttt{qft.line}, even though the imported object no longer has the required type. This unreported loss is the synthesis availability problem studied in this paper.



\section{Background}\label{sec:background}


\subsection{Quantum Compilation and High-Level Synthesis}
\label{sec:bg:hls}

Qiskit calls the synthesis of high-level operations \emph{high-level synthesis} (HLS)~\cite{qiskithls}. A high-level operation may have multiple synthesis methods with different costs and resource requirements. The applicable methods can depend on target connectivity and available ancillas. Qiskit, for example, provides MCX synthesis methods that require no ancillas and other MCX synthesis methods that use clean ancillas~\cite{qiskithls}. \emph{Clean ancillas} are spare qubits known to start in a fixed state, whereas \emph{dirty ancillas} have no fixed initial state. Constructions using clean ancillas can substantially reduce gate counts~\cite{khattar2024}. Linear reversible functions can similarly be synthesized using Gaussian elimination~\cite{pmh2008} or nearest-neighbor methods~\cite{kms2007}. These synthesis methods remain available only while the compiler retains a representation on which they can act.

\subsection{Representation Changes and Routes}\label{sec:bg:repr}

We distinguish four forms of representation change. \emph{Framework serialization}, such as QPY or the JSON formats of Cirq and TKET, stores and restores framework-specific circuit objects~\cite{cirq,qiskitqpy,pytketdocs}. \emph{OpenQASM interchange} represents the program as text that is parsed again by the receiving compiler~\cite{openqasm2,openqasm3paper}. \emph{Compiler conversion} maps operations into the receiving compiler's operation vocabulary~\cite{pytketqiskit}. \emph{Explicit lowering} replaces high-level operations with lower-level gates~\cite{qiskit2024}.

We call each concrete procedure used to change a circuit's representation a \emph{representation route} (or simply \emph{route}). A representation route \emph{delivers} when it produces a circuit for downstream compilation, the \emph{delivered circuit}; if it produces none, we record a \emph{delivery failure}. A \emph{direct OpenQASM~3 round trip} exports a circuit to OpenQASM~3 and immediately imports it back, without prior lowering or conversion through another compiler.

\subsection{Functional Equivalence and Synthesis Availability}
\label{sec:bg:equiv}

Two unitary circuits are \emph{functionally equivalent} when they implement the same unitary transformation up to a global phase. Existing approaches can establish functional equivalence using decision diagrams~\cite{qcec2021}, ZX-calculus rewriting~\cite{peham2022zx}, or symbolic path sums~\cite{amy2019feynman}. We use \emph{functional equivalence} for this property and \emph{operator-equivalence checking} for the matrix-based procedure used in our experiments.

A delivered circuit has \emph{synthesis availability} for a high-level operation when the receiving compiler can apply a synthesis method directly to that operation in the delivered representation, without first reconstructing the high-level operation from lower-level gates. Synthesis availability is therefore relative to the receiving compiler and its version. A \emph{synthesis request} names a synthesis method explicitly. We distinguish three outcomes: the requested method is not invoked; it is invoked but returns no synthesis circuit; or it returns a synthesis circuit. Only the third outcome shows that the request takes effect. Functional equivalence does not determine synthesis availability: functionally equivalent circuits can differ in whether synthesis remains available for the corresponding high-level operation. Qiskit provides synthesis through HLS dispatch, TKET through its native high-level vocabulary, and Cirq through gate decomposition~\cite{cirq,qiskithls,tket2020}.

\subsection{Operation Records and Record Validity}\label{sec:bg:records}

An \emph{operation record} stores information needed to reconstruct a high-level operation, including its operation type, parameters, and qubit binding. OpenQASM~3 directives provide one possible carrier for such information~\cite{openqasmspec}. Classical compilers face a related validity problem: LLVM specifies when retained analysis results or transformation metadata become invalid~\cite{llvmnewpm,llvmtransformmeta}.

An operation record is \emph{valid} when the intended circuit region remains identifiable and the high-level operation reconstructed from the record is functionally equivalent to that region; otherwise, the record is \emph{stale}. Later compiler transformations can make a record stale: routing can change the qubit binding~\cite{li2019sabre,qiskittranspileparams}, optimization can rewrite or merge circuit regions~\cite{qiskit25release}, and approximate compilation can change the region's computation~\cite{qiskittranspileparams}. None of the four forms of representation change (see Section~\ref{sec:bg:repr}) tracks record validity~\cite{cirq,openqasm2,tket2020,openqasm3paper,openqasmspec,pytketqiskit,qiskitqpy}, so a record can remain well formed after it becomes stale.

\section{Empirical Study}\label{sec:study}

To answer RQ1 (which representation changes preserve synthesis availability, and when does losing it affect compiled circuit cost?) and RQ2 (can a retained operation record restore synthesis availability lost at a representation boundary while preserving functional equivalence, and how much time does verification require?), we design the experiments described in this section.


In the controlled experiments, a \emph{condition} fixes the operation and compilation settings while the route, reference, or recovery policy varies. For MCX, a condition is defined by the number of controls, target connectivity, available ancillas, and requested synthesis method.

\subsection{Subjects and Experimental Setup}\label{sec:study:subjects}
\emph{High-level operations and instances.}
We study four Qiskit high-level operation families: quantum Fourier transform (QFT), multi-controlled X (MCX), \texttt{LinearFunction}, and \texttt{PauliEvolutionGate} (PauliEvolution). We select these families because each has more than one synthesis method in Qiskit~2.5.2~\cite{qiskithls}, so the choice of method can change the compiled circuit. The compilation references (see Section~\ref{sec:design:conditions}) and Qiskit representation-route study (see Section~\ref{sec:design:boundary}) use 12 \emph{study-designed instances}: QFT on $n \in \{6,10,14,18\}$ qubits; MCX with $\{4,6,8,10\}$ controls and two clean ancillas; and \texttt{LinearFunction} on $n \in \{6,10,14,18\}$ qubits, with invertible binary matrices generated from a fixed seed.

\emph{Compilers.}
We use Qiskit~2.5.2~\cite{qiskit2024,qiskithls}, pytket~2.18.1~\cite{pytketdocs,tket2020} with pytket-qiskit~0.78.0~\cite{pytketqiskit}, and Cirq~1.6.1~\cite{cirq}. The Munich Quantum Toolkit (MQT) pipeline uses MQT Bench~2.3.0~\cite{quetschlich2023mqtbench}, MQT ProblemSolver at the pinned revision recorded in the replication package~\cite{sweReplicationPackage2026}, and the Microsoft Quantum Resource Estimator in the Quantum Development Kit (QDK)~1.32.3~\cite{vandam2023resourceestimator}.

\emph{Compilation targets, metrics, and repetition.}
\label{sec:study:analysis}\label{sec:design:analysis}
Unless stated otherwise, Qiskit cost comparisons compile each circuit at optimization level~3 with SABRE layout and routing~\cite{li2019sabre}, the basis $\{\texttt{rz},\texttt{sx},\texttt{x},\texttt{cx}\}$, and line, approximately square grid, and heavy-hex connectivities~\cite{chamberland2020}. Each target has as many qubits as the circuit, including ancillas; heavy-hex uses a connected patch of a distance-3 lattice. Because SABRE is randomized, each condition uses ten routing seeds. The primary cost measure is the median two-qubit gate count~\cite{murali2019noise}, \emph{two-qubit count} for short; secondary measures are two-qubit depth and, for the compilation references, total circuit depth. When a QPY round trip is the comparator, we call it the \emph{QPY control}.

\emph{Operator-equivalence checking.}
We remove barriers and construct dense operator matrices. For compiled circuits, we apply the stored final layout so that routing permutations are included. We normalize global phase using the largest-magnitude entry and compare entries with absolute and relative tolerances of $10^{-8}$. For MCX with clean ancillas, we compare only inputs in which those ancillas start in zero. Because dense operators scale as $O(4^n)$, this check is limited to ten qubits unless a study explicitly defines a separate check for wider circuits.

\subsection{RQ1: Synthesis Availability and Compilation Consequences}
\label{sec:study:rq1}
We answer RQ1 with five studies. The first study measures how synthesis choice changes cost.
\subsubsection{Compilation references and optimization opportunity}\label{sec:design:conditions}
We compare three compilation references. The \emph{early-synthesis reference} applies Qiskit's default synthesis at optimization level~0, treating idle qubits as clean ancillas, and translates the result to the basis gates before the measured downstream compilation. The \emph{default-synthesis reference} compiles the intact high-level operation using Qiskit's default synthesis policy, which normally selects the first applicable method~\cite{qiskithls}. The \emph{best-observed reference} compiles the intact operation with every documented applicable synthesis method and the default policy, then keeps the result with the lowest median two-qubit gate count, breaking ties by two-qubit depth, total depth, and method name.

The 12 study-designed instances and three connectivities give 36 conditions. A \emph{resource sweep} applies the same protocol to 102 conditions that vary family, size, clean-ancilla availability, and connectivity. A condition has an \emph{optimization opportunity} when the best-observed reference uses fewer two-qubit gates than the early-synthesis reference.

\subsubsection{Representation-route study}\label{sec:design:boundary}
We evaluate synthesis availability (see Section~\ref{sec:bg:equiv}) across representation routes in Qiskit, TKET, and Cirq.

\emph{Qiskit and TKET.}
The Qiskit study evaluates each study-designed instance after a QPY round trip, the \emph{OpenQASM~3 delivery procedure}, and a Qiskit--TKET--Qiskit round trip. The delivery procedure first attempts a direct OpenQASM~3 round trip; if export or import fails, it lowers the circuit by one additional decomposition level and retries, up to ten levels, retaining the first successful round trip or recording a delivery failure. After the Qiskit--TKET--Qiskit round trip, availability is measured in Qiskit; TKET as the receiving compiler is studied separately.

The TKET study uses eight instances: four MCX and four \texttt{LinearFunction}. QFT is excluded because pytket~2.18.1 has no comparable high-level QFT operation~\cite{pytketdocs}. We compare native construction (the reference), a JSON round trip, OpenQASM~2 text, and direct conversion from Qiskit.

\emph{Cirq.}
The Cirq study uses \texttt{QuantumFourierTransformGate}, controlled-X \texttt{ControlledGate}, and \texttt{PauliStringPhasorGate} at widths $w \in \{4,6,8,10\}$; exact gate parameters are provided in the replication package~\cite{sweReplicationPackage2026}. We compare native construction (the reference), a JSON round trip, an OpenQASM~3 round trip through Cirq's own interface, and explicit lowering. Each delivered circuit is compiled with \texttt{cirq.optimize\_for\_target\_gateset} onto \texttt{CZTargetGateset}~\cite{cirq}. We record whether each decomposition method runs during the representation route or only during downstream compilation, check functional equivalence to the reference after both stages, report two-qubit count and moment depth (the number of Cirq moments), and repeat every compilation without instrumentation to check for instrumentation effects.

We label a delivered circuit \emph{High-level} when a recognized high-level operation remains, \emph{Name only} when only its name remains on a generic object, and \emph{Lowered} when only lower-level operations remain.

\emph{Cross-version check.}
For one QFT, one MCX, and one \texttt{LinearFunction} instance, we generate QPY and direct OpenQASM~3 artifacts with Qiskit~1.0.2, 1.4.6, and 2.0.3 and read them with Qiskit~2.5.2. Thirteen of the 18 planned combinations produce an artifact: Qiskit~1.0.2 lacks the \texttt{QFTGate} used here, and none of the three releases can export \texttt{LinearFunction} to OpenQASM~3.

\subsubsection{Project study}\label{sec:design:corpus}
Our \emph{corpus} is a pinned snapshot of independently authored projects in the Qiskit Ecosystem catalog.\footnote{\url{https://github.com/Qiskit/ecosystem}, cloned 2026-08-24; the catalog commit is recorded in the replication package~\cite{sweReplicationPackage2026}.} Each call site that constructs a studied operation with literal arguments yields an \emph{extracted instance}. Deprecated QFT blueprint constructions are normalized to a functionally equivalent \texttt{QFTGate} (36 of 44 QFT sites; the other eight already use \texttt{QFTGate}), and \texttt{QuantumCircuit.mcx} calls are normalized to \texttt{MCXGate}. Each distinct extracted instance is evaluated after QPY, the OpenQASM~3 delivery procedure, and a Qiskit--TKET--Qiskit round trip. Table~\ref{tab:corpus} traces the reduction from 179 catalog entries to eight extracted instances; one acts on a single qubit, where connectivity does not matter, so cost comparisons use the other seven.

\begin{table}[htbp]
\caption{Qiskit Ecosystem corpus reduction, runtime confirmation, and project code paths}
\label{tab:corpus}
\footnotesize
\begin{tabular}{@{}lrl@{}}
\toprule
Stage & Count & Detail \\
\midrule

\multicolumn{3}{@{}l@{}}{\emph{Extracting operation instances from source}} \\

Catalog entries & 179 &
178 distinct repositories at pinned commit \\

Scanned catalog entries & 155 &
154 repositories; 24 non-Python entries excluded \\

Repositories using the operations & 29 &
MCX 21, QFT 11, \texttt{LinearFunction} 3; counts overlap \\

Call sites & 148 &
excludes 47 import-only references \\

Literal-argument call sites & 38 &
the other 110 have non-literal arguments \\

Distinct extracted instances & 11 &
keyed by family and full literal parameters \\

Extracted instances used & 8 &
from 9 repositories; 2 lack specialized synthesis, 1 too wide \\

Cost conditions & 21 &
7 instances $\times$ 3 connectivities \\

\midrule \multicolumn{3}{@{}l@{}}{\emph{Running repositories that use the studied operations}} \\

Initial runtime pass & 20/29 &
executed; 1 repository reached a studied operation \\

Captured-circuit study & 21/29 &
executed; 7 reached an operation; 5 outside benchmark/serializer tests \\

\midrule \multicolumn{3}{@{}l@{}}{\emph{Representation changes and project code paths}} \\

Representation-change call sites & 394 &
QPY, OpenQASM, conversion, and lowering \\

Validated code paths & 51 &
across 13 repositories; QPY: 0, OpenQASM~2: 15, OpenQASM~3: 1, \\
& & conversion: 21, lowering: 14 \\

Dynamically confirmed code paths & 25 &
across 2 of the 13 repositories, under project tests \\

\bottomrule
\end{tabular}
\end{table}

\emph{Width characterization.}
We also resolve constructor values at 17 of the 110 non-literal sites from local constants, default values, callers, tests, and documented examples, only to characterize widths. Widths are then known for 55 of 148 call sites (the 38 literal sites plus these 17): 54 have 1--8 qubits, and one is a 50-qubit QFT. All but the 50-qubit QFT thus fall within the ten-qubit limit of operator-equivalence checking; the unresolved sites prevent a corpus-wide coverage estimate.

\emph{Captured circuits.}
We capture circuits while running the projects' tests. For the main analysis, we retain circuits containing an MCX, QFT, or \texttt{LinearFunction} operation at top level. Capture uses two passes over the 29 repositories and continues past individual test failures; the second follows a frozen addendum to the declared test dependencies (coverage in Table~\ref{tab:corpus}). This yields 259 captured circuits, including 216 with an MCX. We compile each on the line target with the default-synthesis reference and ten routing seeds after four representation variants: the QPY control, an OpenQASM~2 round trip, a direct OpenQASM~3 round trip, and the OpenQASM~3 delivery procedure.

The cost analysis concerns the MCX subset. A qubit outside an MCX counts as a clean ancilla when no dependency-graph predecessor of the MCX other than a barrier or delay acts on it. The 200 MCX circuits with established functional equivalence enter the cost analysis: 125 with a clean ancilla available at an MCX and 75 without. The two 11-qubit circuits are checked column by column on inputs in which the classified clean ancillas start in zero.

\emph{Arithmetic circuits and additional operation families.}
Under a protocol frozen before any outcome was inspected, we also analyze the captured circuits outside the 259: 152 without a top-level MCX, QFT, or \texttt{LinearFunction}, and 50 that the QPY reader of Qiskit~2.5.2 cannot restore. Of these 50, excluding three \texttt{MCPhaseGate} circuits and 23 copies of one MQT Bench compiler intermediate leaves 24 captured circuits of adder and multiplier gates from MQT Bench and Qamomile~\cite{qamomile}, which form 17 distinct \emph{arithmetic circuits}. We read these with a patched QPY reader that must reproduce each recorded hash, and we use the compilation with no representation change as their reference. Every arithmetic circuit that increases by at least 10\% is also regenerated from its project's pinned source. We also compile 37 study-designed instances from all 14 operation families with registered synthesis methods in Qiskit~2.5.2 in the same way, without and with four clean ancillas.
 
\emph{Project audit.}
We manually inspect candidate representation-change call sites and count a \emph{code path} (Table~\ref{tab:corpus}) when the resulting circuit is later transpiled, executed, converted, serialized, or decomposed along the same source data flow. A code path is \emph{dynamically confirmed} when that downstream use occurs in the project's tests or documented execution path. We also trace each non-default synthesis selection, made through \texttt{HLSConfig} or during PauliEvolution construction, to the first compilation or representation change reached by the circuit, including decomposition inside the construction helper. A \emph{rebuild site} is project code that rebuilds a high-level operation from the current circuit before that operation reaches synthesis along the same data flow.

\subsubsection{Request study}\label{sec:design:requests}

We make each synthesis request through \texttt{HLSConfig}~\cite{qiskithls}, first on the QPY control and then after each representation route. When a route delivers a circuit, a request has \emph{no effect} if the requested method is not invoked or returns no synthesis circuit. The OpenQASM~3 comparisons use the direct OpenQASM~3 round trip, not the delivery procedure. For QFT, MCX, and \texttt{LinearFunction}, we request every documented synthesis method in Qiskit~2.5.2: 17 registered identifiers naming 16 distinct methods.

\emph{Population.}
QFT and MCX use the extracted instances (see Section~\ref{sec:design:corpus}); \texttt{LinearFunction} uses study-designed instances because no extracted \texttt{LinearFunction} instance meets the request-study selection criteria. For MCX, the three- and four-control instances each use 0, 1, or 2 clean ancillas, and the five-control instance also uses 3, giving ten instance--resource settings. We exclude four settings intended to provide dirty ancillas because Qiskit classifies those qubits as clean. Crossing the retained settings with the requested methods and three connectivities gives 411 candidate conditions: 330 MCX, 45 QFT, and 36 \texttt{LinearFunction}. A candidate is \emph{eligible} when the requested method is invoked on the QPY control and returns a synthesis circuit. Ninety-three MCX candidates fail this criterion because the method's resource requirements are unmet, and nine one-qubit QFT candidates have no target connectivity, leaving 309 eligible conditions.

\emph{PauliEvolution instances.}
PauliEvolution uses 60 deterministic instances of mutually commuting Pauli terms at widths 5--10. At each width, the instances cover line, ring, star, ladder, and mixed-locality structures in diagonal and general commuting forms; mixed locality combines Pauli strings of weights one, two, and three, and the general forms are obtained by deterministic Clifford conjugation. Because the terms commute, the first-order product formula introduces no approximation error, so the same functional-equivalence criterion applies. We request the Rustiq synthesis method with two-qubit-count optimization~\cite{goubaultdebrugiere2024faster,qiskithls}; the 60 instances and three connectivities give 180 eligible conditions.

\emph{Dispatch instrumentation and source audit.}
We instrument Qiskit's HLS dispatch and record \emph{invoked}, when the stage calls the requested method, and \emph{produced}, when that method returns a synthesis circuit. Instrumented and uninstrumented controls agree in structure and compilation metrics for 93 sampled QFT, MCX, and \texttt{LinearFunction} cases and 144 paired PauliEvolution compilations. We also inspect all 55 registered synthesis entry points~\cite{qiskithls} and classify whether applicability depends on the delivered object's type or operation-specific attributes.

\emph{Cost comparisons and beneficial requests.}
A cost comparison pairs a representation variant with the QPY control for the same condition and requires both seed-0 compilations to be functionally equivalent to the original operation; this also limits QFT, MCX, and \texttt{LinearFunction} to ten qubits. The filters leave 291 pairs each for no representation change, the Qiskit--TKET--Qiskit round trip, and the early-synthesis reference, and 180 for the direct OpenQASM~3 round trip. All 180 PauliEvolution conditions enter the QPY and direct-OpenQASM~3 comparisons. A request is \emph{beneficial} on a routing seed when the requested method uses fewer two-qubit gates than the default-synthesis reference on the QPY control, and at the condition level when its ten-seed median is lower.

\subsubsection{End-to-end MQT pipeline}\label{sec:design:mqt}
MQT ProblemSolver~\cite{quetschlich2023mqtproblemsolver} accepts MQT Bench circuits~\cite{quetschlich2023mqtbench} or circuits loaded from OpenQASM~2 and sends both through the same Qiskit compilation and Microsoft Quantum Resource Estimator~\cite{vandam2023resourceestimator}. In \emph{Direct}, the compilation receives Bench's target-independent Grover circuit~\cite{grover1996} on six or eight qubits, generated at Bench's default optimization level~2; in \emph{Via OpenQASM~2}, the same circuit is exported to OpenQASM~2 and reloaded. No synthesis request is made: both use ProblemSolver's normal downstream compilation at optimization level~1 with the basis of its resource-estimation pipeline. We check functional equivalence, record whether MCX synthesis produces a circuit, and measure two-qubit count and depth, the physical-qubit estimate, and estimated fault-tolerant runtime. The main comparison leaves the transpiler seed unset; seeds~0--9 test robustness.

\subsection{RQ2: Verified Recovery of Synthesis Availability}\label{sec:study:rq2}\label{sec:design:pipelines}
We answer RQ2 in Qiskit, where the synthesis requests are made (see Section~\ref{sec:design:requests}).
\subsubsection{Operation record and recovery policies}
\label{sec:study:policies}

\Needspace*{11\baselineskip}%
\begin{wrapfloat}{listing}{r}{0.53\columnwidth}
\caption{Example operation record for a five-qubit QFT}
\label{lst:record}
\lsttop
\begin{lstlisting}[style=scq, numbers=none, xleftmargin=2pt]
{"name": "qft",
 "descriptor": {"kind": "qft",
    "parameters": {"inverse": false}},
 "binding": {"qubit_map": [0,1,2,3,4],
    "fingerprint": "<fingerprint>"}}
\end{lstlisting}
\lstbot
\end{wrapfloat}

We keep each \emph{operation record} in a separate file that later transformations do not update (Listing~\ref{lst:record}). Its \texttt{name} is used to find candidate regions, its descriptor stores the operation kind and parameters, and its binding stores the recorded \texttt{qubit\_map} and a fingerprint of the region. A \emph{standalone circuit} contains only the recorded region, whereas an \emph{embedded circuit} also contains surrounding operations.

We compare three recovery policies after a representation boundary. \emph{Drop} discards the record and compiles the delivered circuit unchanged. \emph{Unchecked}, used on standalone circuits, replaces the delivered circuit with the operation reconstructed from the record on the recorded \texttt{qubit\_map}, without verification. \emph{Verified} checks the record against the delivered circuit before reconstruction and places the operation on the delivered region's qubits.

For the main stale-record set (see Section~\ref{sec:study:groundtruth}), we also evaluate two simpler checks. \emph{Name check} accepts when exactly one delivered operation carries the recorded name. \emph{Descriptor check} accepts when the descriptor reconstructs an operation accepted by the family's synthesis method. Neither shows that the reconstructed operation is functionally equivalent to the delivered region. Verified has three stages: \emph{Locate}, \emph{Verify}, and \emph{Reconstruct}.

\begin{enumerate}[nosep,leftmargin=1.6em]
\item \emph{Locate.} Find the delivered region to check. Unless the experiment supplies the region directly, Locate searches for the delivered operation carrying the recorded name and rejects the record when the candidate is missing or ambiguous. If the region lies inside a nested definition, Locate descends into it and checks each recorded occurrence separately.

\item \emph{Verify.} Accept on a matching fingerprint. Otherwise, perform a bounded operator-equivalence check between the operation reconstructed from the descriptor and the located region in its delivered qubit order, without applying the circuit's stored final layout. Regions larger than ten qubits are accepted only on a matching fingerprint.

\item \emph{Reconstruct.} Reconstruct the high-level operation from the descriptor and place it on the verified region's qubits in their delivered order. For a region inside a nested definition, rebuild that definition with only the verified region replaced, leaving the enclosing operation unchanged.
\end{enumerate}

\emph{Fingerprint check.} The fingerprint is the SHA-256 digest of the region's QPY serialization~\cite{qiskitqpy} after barriers are removed, taken when the record is created. A match provides a fast check that the recorded representation is unchanged. A representation change can alter the fingerprint without making the record stale, so Verified falls back to operator-equivalence checking after a mismatch.

\emph{Assumptions and limits.} Verified assumes that the record is correct when created, that the supplied or located region and its qubit order are correct, that the fingerprint does not collide, and that the bounded equivalence check is correct. A matching fingerprint shows only that the recorded representation is unchanged; it does not independently validate the original record. 


\subsubsection{Recovery with valid records}\label{sec:study:recovery}

The \emph{recovery population} has 83 conditions: the eight beneficial MCX conditions retained by the delivery and equivalence filters (see Section~\ref{sec:design:requests}), all 36 eligible QFT conditions, the 21 PauliEvolution conditions in which Rustiq is beneficial on the QPY control, and 18 eligible \texttt{LinearFunction} conditions within the ten-qubit equivalence limit. In every included condition, the requested method produces a synthesis circuit before the representation change; afterward, the route delivers a functionally equivalent circuit without synthesis availability. Drop therefore cannot restore the request.

MCX, QFT, and PauliEvolution are recovered after a direct OpenQASM~3 round trip. Because the direct OpenQASM~3 round trip does not deliver \texttt{LinearFunction} in this population, those conditions use a Qiskit--TKET--Qiskit round trip; the operation fills the whole circuit and its name does not survive conversion, so the delivered region is supplied directly. The three stages Locate, Verify, and Reconstruct are otherwise the same for all families.

\emph{Original} compiles the high-level representation directly with the same requested method used for Drop and Verified. Ten routing seeds give 830 runs each for Original, Drop, and Verified, 2,490 in total. A request is counted as restored only when the requested method is invoked on the reconstructed operation and produces a synthesis circuit. Each run also records two-qubit count and depth, functional equivalence, verification outcome, and verification time.

Before the OpenQASM~2 round trip in the MQT pipeline, we create a record for each MCX and \texttt{MCPhaseGate} region inside the Grover operator definition. The recorded \texttt{qubit\_map} uses the definition's local qubit indices and is rebound to the enclosing operation's qubits after verification. Qiskit~2.5.2 has no registered synthesis method for \texttt{MCPhaseGate}~\cite{qiskithls}, so reconstructing only those regions tests whether rebuilding the nested definition changes the result without restoring MCX synthesis availability. Starting from Via OpenQASM~2, we reconstruct MCX alone, \texttt{MCPhaseGate} alone, and both, using identical downstream compilation and resource-estimation settings. 

\subsubsection{Stale records and resynthesis}
\label{sec:study:workflows}\label{sec:study:groundtruth}

To test recovery when operation records may become stale, we apply six transformation sequences without updating the record: a QPY round trip~\cite{qiskitqpy}; a direct OpenQASM~3 round trip~\cite{openqasm3paper}; optimization level~2 to the basis gate set; optimization level~3 with SABRE routing~\cite{li2019sabre} to a linear connectivity followed by OpenQASM~3; conversion to TKET with full peephole optimization~\cite{pytketdocs} and conversion back~\cite{pytketqiskit}; and optimization level~3 with approximation degree~0.9.

Each transformation sequence is applied to six circuits; one circuit--sequence pairing is one \emph{case}. Four standalone circuits contain only the recorded operation: QFT on 5 and 8 qubits, MCX with four controls, and \texttt{LinearFunction} on 6 qubits. Two embedded circuits place the recorded operation among surrounding operations: a five-qubit Qiskit Grover operator containing an MCX~\cite{grover1996}, and a six-qubit Greenberger--Horne--Zeilinger circuit~\cite{ghz1989}. For the latter, we apply Qiskit's \texttt{CollectLinearFunctions} pass to group the controlled-X chain into a \texttt{LinearFunction}. We evaluate record acceptance separately from reconstruction correctness; for every accepted record, we compare the reconstructed circuit with the delivered circuit using operator-equivalence checking. For the four standalone circuits, the region is supplied directly.

\emph{Independent validity labels.}
To label record validity independently of the recovery policies, we use a separate implementation of operator-equivalence checking (see Section~\ref{sec:study:subjects}) and require agreement with the MQT Quantum Circuit Equivalence Checking (QCEC) tool, version~3.10.0~\cite{qcec2021}; when QCEC's default configuration gives no result, we use its decision-diagram checker. For standalone circuits, the candidate region is the whole delivered circuit. For embedded circuits, we search contiguous sequences of delivered operations without using the recorded name.

The 36 cases above form the \emph{main stale-record set}; 28 of them deliver a circuit. The \emph{extended stale-record set} applies the five sequences other than the TKET one to 11 standalone circuits of 3--14 qubits, giving 55 cases; 49 deliver a circuit and six fail OpenQASM~3 import or export. Operator-equivalence checking is available for 41 of the 49 delivered cases.

The \emph{resynthesis set} tests whether Verified still accepts valid records after resynthesis. It contains QFT at widths 6 and 10, MCX with 4 and 6 controls, and \texttt{LinearFunction} at widths 6 and 10; the default synthesis policy and every applicable documented method give 30 cases. Descriptor check is also evaluated on the resynthesis set.

\subsubsection{Verification timing}\label{sec:study:timing}

Verification time is wall-clock time and includes fingerprinting and, after a mismatch, reconstruction for operator-equivalence checking and the check itself; it excludes Locate, placement, and compilation. Each fingerprint check is timed five times per condition, while recovery and stale-record verification are timed once per run. Verification and compilation times are compared only within the same run. We also time the fingerprint check on the eight MCX recovery conditions for routes that leave the recorded fingerprint unchanged.
\section{Results}\label{sec:results}
We report the results for RQ1 and then for RQ2.

\subsection{RQ1: Synthesis Availability and Compilation Consequences}
\label{sec:rq1}
In short, synthesis availability survives a representation change only when the high-level operation survives, and the evaluated OpenQASM routes remove the operation. After a direct OpenQASM~3 round trip, a synthesis request has no effect, yet compilation succeeds. The two-qubit count then rises, falls, or stays the same, depending on the synthesis that is lost.
\subsubsection{Synthesis availability depends on representation and receiving compiler}\label{sec:rq1:boundaries}
\begin{figure}[!t]
\centering
\includegraphics[width=\linewidth]{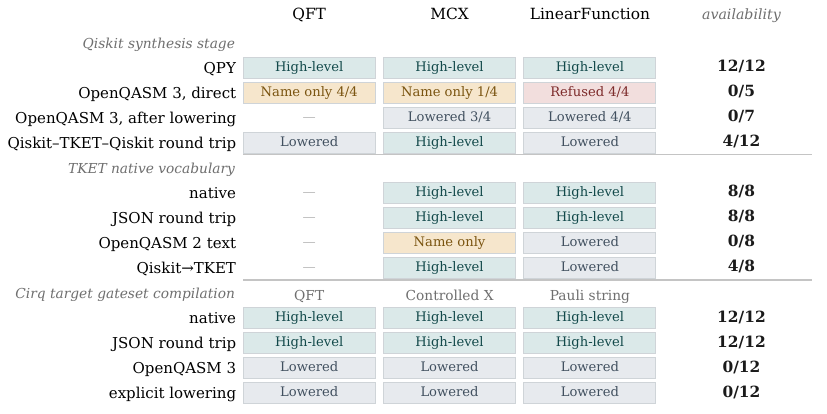}
\caption{Synthesis availability in the representation-route study. For cell labels, see Section~\ref{sec:design:boundary}; within-family fractions use the attempted instances, the rightmost availability denominator counts only instances delivered at that step, and \emph{Refused} marks delivery failures. Each Cirq cell summarizes widths 4, 6, 8, and 10 qubits, which have identical outcomes.}
\label{fig:availability}
\end{figure}

Figure~\ref{fig:availability} summarizes the representation-route study. In Qiskit, QPY preserves synthesis availability for all 12 study-designed instances, the OpenQASM~3 delivery procedure for none, and the Qiskit--TKET--Qiskit round trip only for the four MCX instances, which TKET represents natively~\cite{pytketdocs}. Five instances survive the direct OpenQASM~3 round trip as generic \texttt{Gate} objects; for the other seven, the delivery procedure must first lower the circuit to gates. With TKET as the receiving compiler, native construction and JSON serialization preserve availability for all eight instances, OpenQASM~2 text~\cite{openqasm2} for none, and direct conversion from Qiskit only for the four MCX instances.

In Cirq, native construction and JSON serialization preserve availability in all 12 conditions, whereas an OpenQASM~3 round trip and explicit lowering remove it in all 12 because both decompose the operations before target gateset compilation. All 96 equivalence checks pass, with a maximum deviation of $5.8\times10^{-13}$. Two-qubit count is identical across routes, although moment depth differs: OpenQASM~3 is 2--96 moments deeper in 8 of 12 conditions, while explicit lowering produces both increases and decreases.

The eight extracted instances (see Section~\ref{sec:design:corpus}) show the same pattern: QPY preserves availability in 8 of 8, the OpenQASM~3 delivery procedure in none, and the Qiskit--TKET--Qiskit round trip in the three MCX instances. The direct OpenQASM~3 round trip delivers seven of the eight instances, while the five-control MCX is delivered only after its circuit is lowered. In the cross-version check, all 13 combinations that produce an artifact show the same availability pattern.

\subsubsection{Synthesis requests have no effect after a direct OpenQASM~3 round trip}\label{sec:rq1:requests}

Of the 489 eligible conditions, the direct OpenQASM~3 round trip delivers 360: 144 MCX, 36 QFT, and all 180 PauliEvolution conditions (Table~\ref{tab:requests}). In every delivered condition, Qiskit invokes the requested method, which receives a generic \texttt{Gate} and returns no synthesis circuit. Because Qiskit's HLS interface permits a method to return no circuit for an unsupported operation~\cite{qiskithls}, compilation continues from the imported gate definition and succeeds for all ten routing seeds in all 360 conditions. None of the 180 delivered MCX and QFT conditions raises an exception or Python warning, and five representative checks emit nothing to standard output, standard error, or Qiskit's log.

\begin{table}[!t]
\caption{Outcome of synthesis requests by representation route or reference}
\label{tab:requests}
\footnotesize
\setlength{\tabcolsep}{4pt}
\begin{tabular}{@{}lrrrrr@{}}
\toprule
Route or reference & Eligible & Produced & Invoked, no circuit & Not invoked & Delivery failure \\
\midrule
\multicolumn{6}{@{}l}{\textit{QFT, MCX, and \texttt{LinearFunction}}} \\
No change or QPY round trip & 309 & 309 & 0 & 0 & 0 \\
OpenQASM~3, direct       & 309 &   0 & 180 & 0 & 129 \\
Qiskit--TKET--Qiskit     & 309 & 237 & 0 & 72 & 0 \\
Early-synthesis reference & 309 &   0 & 0 & 309 & 0 \\
\midrule
\multicolumn{6}{@{}l}{\textit{PauliEvolution}} \\
No change or QPY round trip & 180 & 180 & 0 & 0 & 0 \\
OpenQASM~3, direct       & 180 &   0 & 180 & 0 & 0 \\
\bottomrule
\end{tabular}
\end{table}

The Qiskit--TKET--Qiskit round trip produces the requested synthesis only when the corresponding high-level operation survives conversion, while the early-synthesis reference leaves no high-level operation for downstream synthesis. The 129 delivery failures of the direct OpenQASM~3 round trip are 36 \texttt{LinearFunction} export failures and 93 five-control MCX import failures.

The source audit shows that type-based applicability is not specific to QFT. Of 55 registered synthesis entry points across 14 operation families, 52 check the delivered object's type and return a circuit only for the expected type; QFT and MCX explicitly reject custom operations that merely share the high-level operation's name~\cite{qiskithls}. The remaining three inspect operation-specific attributes and raise an exception for incompatible objects, so they report the mismatch instead of silently returning no circuit. We observe this behavior for IntegerComparator, whose synthesis raises an exception after both OpenQASM~3 routes.

The project audit finds 64 non-default synthesis selections in 11 repositories. Of the 27 made through \texttt{HLSConfig}, 21 reach compilation and none reaches a representation change. Of the 37 made during PauliEvolution construction, 33 cross a representation change, including 27 inside the construction helper. Project code thus contains both non-default synthesis selections and representation changes, but no \texttt{HLSConfig} request in this snapshot crosses a representation change; the evidence that such requests silently lose their effect comes from the request study.

\subsubsection{Cost change depends on the synthesis that is lost}\label{sec:rq1:external}

After a direct OpenQASM~3 round trip, requested and default compilations both start from the imported gate definition, so any cost change depends on how the lost synthesis differs from that definition. Across the 21 cost conditions of the extracted instances, the OpenQASM~3 delivery procedure and QPY control both match the default-synthesis reference in two-qubit gate count. The direct round trip delivers 18 conditions; for the three five-control MCX conditions that are delivered only after lowering, the default-synthesis reference is itself a median of 7.7\% above the best-observed reference. All seven distinct delivered circuits are functionally equivalent to the original operation, and their seed-0 compilations on the line target are functionally equivalent to the default-synthesis ones.

Unlike the extracted instances, the captured MCX circuits include surrounding qubits, and MCX synthesis can use a clean ancilla among them before the representation change but not after; 38 of these circuits rise by at least 10\% (see Table~\ref{tab:mechanisms}). Seven of the 125 MCX circuits with a clean ancilla, all from qclib~\cite{qclib}, already lose MCX synthesis availability in the QPY round trip, which turns their MCX into a generic \texttt{ControlledGate}, so they cannot show a further loss caused by OpenQASM. Thirty-one of the 38 MCX increases come from Qamomile, so this population does not estimate prevalence across projects. The arithmetic increases need no ancilla. These five increases occur in the one-bit full adder of MQT Bench and of Qamomile (from 14 to 22 two-qubit gates) and in the two- to four-bit modular adders of MQT Bench. At these sizes, default synthesis~\cite{qiskithls} selects the ripple-carry adder of Vedral et al.~\cite{vedral1996} and the QFT-based adder of Draper~\cite{draper2000}, respectively, whereas the imported gate definitions implement the ripple-carry adder of Cuccaro et al.~\cite{cuccaro2004} and an ancilla-free modular adder~\cite{gidney2017}. Regeneration from each project's pinned source reproduces all five increases.

\begin{table}[t]
\caption{Two-qubit (2Q) change after OpenQASM routes in captured circuits and study-designed instances}
\label{tab:mechanisms}
\footnotesize
\setlength{\tabcolsep}{4pt}
\begin{tabular}{@{}p{0.23\linewidth}p{0.50\linewidth}p{0.21\linewidth}@{}}
\toprule
\raggedright Evidence & \raggedright 2Q change & \raggedright Condition for a change \tabularnewline
\midrule
\raggedright Captured MCX circuits (200) & \raggedright 38 of the 125 with a clean ancilla rise by 10--164\%; none of the 75 without one rises by 10\% or more & \raggedright Clean ancilla available at an MCX \tabularnewline
\raggedright Arithmetic circuits (14 of 17) & \raggedright 5 rise by 25--57\%; the other 9 change by less than 10\% & \raggedright Adder size (no ancilla) \tabularnewline
\raggedright Other captured (90 of 152) & \raggedright No change; the other 62 do not pass the protocol's filters & \raggedright --- \tabularnewline
\raggedright 37 instances from 14 operation families & \raggedright Rises with four clean ancillas for MCX, MCMT, HalfAdder, and IntegerComparator (23--688\%) and at some sizes for FullAdder and ModularAdder (25--57\%); falls for Multiplier; no change for QFT, \texttt{LinearFunction}, Clifford, Permutation, PauliEvolution, and annotated operations & \raggedright Qiskit's source shows default synthesis differing from the imported gate definition \tabularnewline
\bottomrule
\end{tabular}
\par\addvspace{3pt}
{\footnotesize\raggedright Reference: the QPY control for captured MCX circuits; no representation change for arithmetic circuits and the 37 instances. For the other three arithmetic circuits, all multipliers, functional equivalence is not established. WeightedSum and the two-bit Multiplier are not compared; IntegerComparator is compared only after the OpenQASM~2 route (see Section~\ref{sec:rq1:requests}). \par}
\end{table}


Table~\ref{tab:paired} compares each requested method with its QPY control. Under the direct OpenQASM~3 round trip, two-qubit count is higher in 48 of 180 pairs, lower in 48, and unchanged in 84; median relative changes in both two-qubit count and depth are zero. In every lower-count pair, the requested method uses more two-qubit gates than default synthesis on the QPY control. PauliEvolution shows the same dependence: Rustiq is beneficial on the QPY control in 21 conditions, ties default synthesis in 12, and is more expensive in 147, while the direct round trip removes its synthesis availability in all 180.

\begin{table}[htbp]
\caption{Two-qubit (2Q) outcomes of QFT, MCX, and \texttt{LinearFunction} requests vs.\ the QPY control}
\label{tab:paired}
\footnotesize
\setlength{\tabcolsep}{4pt}
\begin{tabular}{@{}lrrrrrr@{}}
\toprule
 & & \multicolumn{3}{c}{Delivered vs.\ QPY 2Q}
 & Median relative & Lower-2Q pairs \\
\cmidrule(lr){3-5}
Route or reference & $n$ & Higher & Lower & Same
& $\Delta$2Q & request above default \\
\midrule
No representation change & 291 &  0 &   0 & 291 & 0.0\% & -- \\
OpenQASM~3, direct        & 180 & 48 &  48 &  84 & 0.0\% & 48 of 48 \\
Qiskit--TKET--Qiskit      & 291 &  2 &  16 & 273 & 0.0\% & 7 of 16 \\
Early-synthesis reference & 291 & 45 & 148 &  98 & $-7.0\%$ & 148 of 148 \\
\bottomrule
\end{tabular}

\par\addvspace{3pt}
{\footnotesize\raggedright $\Delta$2Q $=$ delivered $-$ QPY control, on ten-seed condition medians; the relative column is the median of $\Delta$2Q$/$QPY across pairs. The last column counts lower-2Q pairs whose request uses more 2Q gates than default synthesis on the QPY control.\par}
\end{table}

\subsubsection{Cost when the requested synthesis reduces two-qubit count}\label{sec:rq1:cost}

Sixteen MCX conditions are beneficial (see Section~\ref{sec:design:requests}). The direct OpenQASM~3 round trip delivers the eight conditions from the three- and four-control instances, and all eight lose the two-qubit reduction, with a median increase of 38.5\% (10.5--152.6\%); for the eight conditions from the five-control instance, the import of the OpenQASM~3 text fails. Across routing seeds, 76 of 79 delivered beneficial MCX comparisons (96.2\%) lose the reduction. No QFT condition (two to five qubits) is beneficial by its ten-seed median, although five routing-seed comparisons are, and all five comparisons lose the reduction.

The 21 beneficial PauliEvolution conditions come from 8 of the 60 operation instances, and Rustiq is cheaper than default synthesis in all 210 of their routing-seed comparisons. All 21 lose the reduction after the direct OpenQASM~3 round trip, with a median increase of 20.0\% (5.0--33.3\%); across all 180 conditions, requested and default compilations have identical two-qubit counts after the round trip in all 1,800 routing-seed comparisons. Of the 21 beneficial conditions, 20 come from mixed-locality instances and one from a ladder instance; line, ring, and star instances contribute none. Seven \texttt{LinearFunction} conditions are also beneficial, but the direct round trip cannot export them, so at the condition level the lost reduction can be measured only for MCX and PauliEvolution.

\subsubsection{Optimization opportunity depends on operation and resources}\label{sec:rq1:envelope}

\begin{figure}[tbp]
\centering
\includegraphics{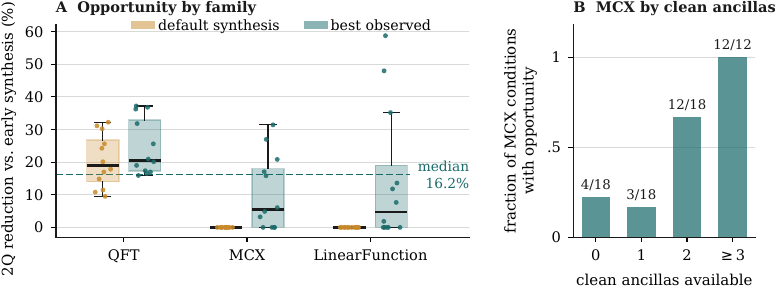}
\caption{Optimization opportunity. (A) Two-qubit reduction relative to the early-synthesis reference; points are the 36 conditions, boxes show quartiles and median, and the dashed line marks the median reduction of the best-observed reference. (B) Fraction of MCX conditions with an opportunity in the resource sweep.}
\label{fig:opportunity}
\end{figure}

Across the 36 conditions of the study-designed instances (see Section~\ref{sec:design:conditions}), the best-observed reference uses a median of 16.2\% fewer two-qubit gates than the early-synthesis reference (Figure~\ref{fig:opportunity}A); it uses fewer in 27 conditions and ties in 9. Median reductions are 18.9\% for two-qubit depth and 7.1\% for total depth. The default-synthesis reference has a median two-qubit reduction of 0.0\% and improves on the early-synthesis reference in all 12 QFT conditions but none of the MCX or \texttt{LinearFunction} conditions.

In the resource sweep, 53 of the 102 conditions have an optimization opportunity; among these 53, the median two-qubit reduction is 15.8\%. The default-synthesis reference is cheaper than the early-synthesis reference in 18 conditions, all QFT. For MCX, the opportunity depends on clean ancillas: 4 of 18 conditions with none, 3 of 18 with one, 12 of 18 with two, and all 12 with at least three (Figure~\ref{fig:opportunity}B). The early-synthesis reference does not model lost availability: it can apply default MCX synthesis with clean ancillas before basis translation, which OpenQASM interchange prevents. Across the 36 conditions, the early-synthesis reference is still cheaper than the best-observed reference in 13 of 360 routing-seed comparisons, all MCX.

\subsubsection{OpenQASM~2 removes MCX synthesis availability in the MQT pipeline}\label{sec:rq1:mqt}

In Direct (see Section~\ref{sec:design:mqt}), downstream compilation reaches four MCX regions in the six-qubit Grover circuit and eight in the eight-qubit circuit. After the OpenQASM~2 round trip, these regions are no longer MCX operations, so MCX synthesis produces no circuit, although Direct and Via OpenQASM~2 remain functionally equivalent. At six qubits, two-qubit count rises from 404 to 440 (+8.9\%), two-qubit depth from 348 to 369, and estimated fault-tolerant runtime by 31.6\%; at eight qubits, two-qubit count rises from 2088 to 2864 (+37.2\%), two-qubit depth from 1681 to 2135, and estimated fault-tolerant runtime by 68.0\%. The physical-qubit estimate is unchanged at both sizes, and the results are unchanged across seeds~0--9. The increases come from the loss of MCX synthesis availability (see Section~\ref{sec:rq2:valid}).

\rqanswer{RQ1}{%
A representation change preserves synthesis availability only when it keeps the high-level operation. The evaluated OpenQASM routes do not: after a direct OpenQASM~3 round trip, requested synthesis has no effect in all 360 delivered conditions. The two-qubit count then changes only where the lost synthesis differs from the imported gate definition.}
\subsection{RQ2: Verified Recovery of Synthesis Availability}\label{sec:rq2}

The project audit finds five rebuild sites in two independently authored repositories. RQ2 asks whether reconstruction from a record is safe.

\subsubsection{Verified recovery restores synthesis availability}\label{sec:rq2:valid}

Across 83 recovery conditions and ten routing seeds, Original and Verified produce the requested synthesis in all 830 runs (Table~\ref{tab:recovery}). By construction, Drop produces it in no run: the requested method returns no synthesis circuit in the 650 direct OpenQASM~3 runs and is never invoked in the 180 \texttt{LinearFunction} runs. All 830 Verified circuits are functionally equivalent to the original program and match Original exactly in two-qubit count and depth. Of the 83 conditions, 31 are beneficial (see Section~\ref{sec:design:requests}): eight MCX, 21 PauliEvolution, and two \texttt{LinearFunction}; none of the 36 QFT conditions is beneficial.

\begin{table}[htbp]
\caption{Valid records accepted and stale records rejected by each recovery policy}
\label{tab:recovery}
{\footnotesize\centering
\setlength{\tabcolsep}{4pt}
\begin{tabular}{@{}lrrl@{}}
\toprule
 & \multicolumn{2}{c}{Verified} & \\
\cmidrule(lr){2-3}
Record set & Valid accepted & Stale rejected & Other policies (valid accepted; stale rejected) \\
\midrule
Recovery population (runs) & 830/830 & -- & Drop: requested synthesis produced in 0/830 \\
Main stale-record set & 13/17 & 11/11 & Name check: 9/17; 11/11. Descriptor check: 17/17; 0/11 \\
Extended stale-record set & 14/24 & 17/17 & Unchecked: reconstructs all 17 stale; all non-equivalent \\
Resynthesis set & 30/30 & -- & Descriptor check: 30/30 \\
\bottomrule
\end{tabular}\par}
\end{table}

In the MQT pipeline, all 24 recorded regions pass Verified's equivalence check. Reconstructing MCX alone returns every measured value to its Direct value. Reconstructing \texttt{MCPhaseGate} alone changes no value, and reconstructing both gives the same values as reconstructing MCX alone. Every reconstructed MQT circuit is functionally equivalent to its Direct circuit.

\subsubsection{Unchecked recovery can violate functional equivalence when records become stale}\label{sec:rq2:stale}

Under the independent validity labels (see Section~\ref{sec:study:groundtruth}), 17 of the 28 delivered cases in the main stale-record set have valid records and 11 have stale records; all six QPY records and all three delivered direct OpenQASM~3 records remain valid. The 11 stale records arise from three mechanisms: approximate recompilation changes the region's unitary in four cases, routing followed by OpenQASM~3 export discards the layout in six cases, and optimization merges the recorded region with surrounding operations in one embedded case. Verified and Name check reject all 11 stale records, whereas Descriptor check accepts them (Table~\ref{tab:recovery}). Verified rejects four valid records because its checks are stricter than the validity labels. In two QFT records, the final swaps exist only in the circuit's final layout; the labels apply this layout, but Verify does not (see Section~\ref{sec:study:policies}). In two embedded \texttt{LinearFunction} records, the region no longer carries the recorded name, so Locate finds no candidate, whereas the labels search without the name.

In the extended stale-record set, 49 cases deliver a circuit and 41 can be independently checked for operator equivalence: 24 with valid and 17 with stale records. The other eight cases involve a 14-qubit \texttt{LinearFunction} or an 11-qubit MCX and exceed the ten-qubit limit. Unchecked reconstructs all 17 stale records and produces a non-equivalent circuit in every case: nine after routing followed by OpenQASM~3 export and eight after approximate recompilation. Verified rejects all 17 stale records, and every circuit it reconstructs in these 41 cases is functionally equivalent to the delivered circuit. Verified also rejects 10 of the 24 valid records because no delivered operation carries the recorded name. In the 30 resynthesis cases, all records remain valid and Verified accepts them.

\subsubsection{Verification coverage and time}\label{sec:rq2:guard}

The recorded fingerprint changes in all 830 recovery runs, all 24 MQT regions, and all 30 resynthesis cases, so Verified uses operator-equivalence checking, which it applies only up to ten qubits. In the recovery runs, median verification time is 2.0~ms for QFT, 6.5~ms for MCX, 107~ms for PauliEvolution, and 638~ms for \texttt{LinearFunction}, with a maximum of 3.8~s. For the eight MCX conditions, the fingerprint matches after no representation change, a QPY round trip, and a Qiskit--TKET--Qiskit round trip (8/8 conditions each). The fingerprint check then takes a median of 0.22, 0.21, and 0.24~ms, respectively; compilation takes a median of 14.5~ms in their recovery runs. Median verification time is 31.5~ms in the resynthesis set (maximum 4.92~s, for a ten-qubit QFT) and 0.49~ms per region for the 49 delivered cases of the extended stale-record set, where compiling ten routing seeds takes a median of 67~ms per case.

\FloatBarrier

\rqanswer{RQ2}{%
Yes, if the record is verified first. Verified restores the requested synthesis in all 830 recovery runs and rejects every stale record, but Verified also rejects some valid records. Unchecked produces a non-equivalent circuit for all 17 stale records of the extended set. Median verification time is 2.0--638~ms per operation family.}
\section{Implications for Quantum Compiler and Interchange Design}
\label{sec:implications}

\emph{Report whether a synthesis request takes effect.}
After a direct OpenQASM~3 round trip, the requested method returns no synthesis circuit in all 360 delivered conditions, yet compilation succeeds (see Section~\ref{sec:rq1:requests}). Compiler developers should make the compiler report whether a requested method is invoked, whether it produces a synthesis circuit, and whether that circuit is applied.

\emph{Preserve high-level operations when later synthesis depends on them.}
Preserving the high-level operation keeps synthesis choices available but does not guarantee a cheaper circuit.
For MCX, the choices can depend on clean ancillas; for the evaluated adders, the choices depend on adder size (see Section~\ref{sec:rq1:external}). When a representation change cannot preserve the high-level operation, developers of interchange tools should retain an operation record for verified recovery.

\emph{Report circuit representation in compiler evaluations.}
Functionally equivalent circuits can differ in synthesis availability and therefore in compiled cost (see Section~\ref{sec:rq1:mqt}). Researchers should report the representation supplied to the compiler along with the workload and compiler settings.

\emph{Verify a retained operation record before reconstruction.}
Unchecked reconstruction changes the computation for all 17 stale records of the extended set (see Section~\ref{sec:rq2:stale}). Developers who reconstruct operations should verify that the record still describes the delivered region and otherwise leave the circuit unchanged, even when this prevents recovery of a valid record.
\section{Threats to Validity}\label{sec:threats}

\emph{External validity.}
To cover different tools and workloads, we study three widely used compiler frameworks, three earlier Qiskit releases, operations and circuits from independently authored Qiskit projects, and an end-to-end MQT pipeline. Our results may still not generalize to other frameworks, releases, or workloads. Recovery with Verified is evaluated only in Qiskit, not in TKET or Cirq. The cross-version check reads artifacts from the earlier releases with Qiskit~2.5.2, so it does not reproduce downstream compilation under those releases. The cost increases come from a few sources. Of the 38 MCX increases, 31 come from one repository. The five arithmetic increases come from small adders in two projects and reflect the implementations selected by Qiskit~2.5.2 at those sizes. Twenty of the 21 beneficial PauliEvolution conditions come from mixed-locality instances, and the MQT pipeline covers one benchmark family at two circuit sizes. We therefore claim only that synthesis availability can be lost and that the loss can affect compiled cost in the evaluated settings; we do not estimate how often this happens elsewhere.

\emph{Construct validity.}
Compiler-specific synthesis interfaces could make synthesis availability difficult to measure consistently. We apply the same definition through each compiler's documented mechanism; these measurements do not imply identical optimization capabilities across frameworks. The commuting PauliEvolution instances introduce no product-formula approximation error, but they do not represent general Hamiltonian simulation. We use two-qubit gate count as the primary cost measure; it does not capture all hardware costs, and the MQT runtime values are resource estimates rather than hardware measurements. Because Qiskit classifies the intended dirty ancillas as clean, we exclude those MCX settings and make no claim about dirty-ancilla synthesis.

\emph{Internal validity.}
Instrumentation could affect compilation. Instrumented and uninstrumented Qiskit controls agree in structure and metrics, and every Cirq condition is also run without instrumentation; the Qiskit controls cover sampled cases rather than every compilation. Selecting requests by their outcome could bias the cost results. We select beneficial requests on the QPY control, before the representation change, and claim lost reductions only for these requests. The patched QPY reader could alter a captured circuit. Every circuit that it restores matches the hash recorded at capture, and every arithmetic increase is reproduced from its project's pinned source.

The circuits captured from repository tests depend on which tests are run. During the first pass, we freeze an addendum to the declared test dependencies. The second pass follows this addendum and supplies 34 of the 38 MCX increases; the two passes remain separate in the experimental record.

Verified is conservative. It rejects every stale record in the evaluated sets but also some valid records. For regions larger than ten qubits, it rejects every record whose fingerprint has changed.

\par\emph{Conclusion validity.}
The evaluated conditions and captured circuits are fixed rather than randomly sampled, and routing seeds measure within-condition variability rather than independent workloads. We therefore report descriptive results without inferential statistics. The best-observed reference is the best result among the tested methods, not a generally optimal synthesis policy.
\section{Related Work}\label{sec:related}

\emph{High-level quantum synthesis and optimization.}
Synthesis methods for the same operation differ in cost under different ancilla and connectivity assumptions~\cite{barenco1995,khattar2024,kms2007,pmh2008}, and Qiskit and TKET offer these alternatives~\cite{qiskithls,tket2020}. Recent work shows that retaining high-level quantum instructions can enable decomposition and compilation optimizations that are unavailable after early lowering~\cite{rosa2025decomp,rosa2025hli}. Paulihedral retains Pauli structure to optimize across Pauli blocks~\cite{paulihedral2022}, while optimization based on assertions uses annotations to enable optimizations across subroutine boundaries~\cite{assertopt2020}. These studies show the benefit of retaining high-level structure; in contrast, we study whether the receiving compiler can still apply synthesis after the circuit changes representation.

\emph{Quantum interchange and semantic preservation.}
OpenQASM~3 permits directives whose interpretation is determined by the receiving compiler~\cite{openqasm3paper,openqasmspec}, while QPY serializes Qiskit circuit objects~\cite{qiskitqpy}. Quantum Intermediate Representation profiles specify functionality supported by a target backend~\cite{qirspec}, and jeff provides an extensible schema for additional features or metadata~\cite{jeffspec}. OpenQASM~3 directives and jeff metadata could carry an operation record; Verified adds a check that the record still matches the delivered circuit.

\emph{Empirical studies of quantum software and compilers.}
In quantum computing platforms, quantum-specific bugs are particularly prevalent in components that represent, compile, and optimize programs~\cite{paltenghi2022bugs}. LintQ analyzes Qiskit applications statically~\cite{lintq2024}. MorphQ tests Qiskit through metamorphic transformations, including OpenQASM round trips~\cite{morphq2023}. QDiff applies differential testing across quantum software stacks~\cite{qdiff2021}, and IterTestQ repeatedly imports, transforms, and exports OpenQASM programs across five platforms under crash and equivalence oracles~\cite{itertestq2026}. Both oracles pass in the cases we study: compilation succeeds and the delivered circuit preserves the computation, yet synthesis is no longer available.

\emph{Equivalence checking and circuit reconstruction.}
Equivalence checkers based on decision diagrams~\cite{qcec2021} or path sums~\cite{amy2019feynman} check that a circuit preserves its computation but do not check whether synthesis remains available. When high-level structure is absent, pattern matching identifies known circuit patterns~\cite{iten2022,qcpm2024}, while decompilation searches for a high-level program consistent with the circuit~\cite{deqompile2025}. Pattern matching and decompilation infer structure from the circuit itself; Verified instead checks an operation record against the delivered circuit before reconstructing the high-level operation, much as translation validation checks a transformed result against its source~\cite{pnueli1998}.
\section{Conclusions}\label{sec:conclusion}

We study synthesis availability across representation changes in Qiskit, TKET, and Cirq. A representation change can preserve a circuit's computation while removing synthesis availability: after a direct OpenQASM~3 round trip, requested synthesis has no effect in all 360 delivered conditions, yet compilation succeeds. The two-qubit gate count can then rise, fall, or remain unchanged, depending on the synthesis that is lost. By checking operation records against the delivered circuit, Verified restores the requested synthesis in all 830 recovery runs and rejects every stale record in the evaluated sets; it also rejects some valid records. Compilers and interchange tools should report whether synthesis requests take effect and verify operation records before reconstruction. In future work, we plan to store operation records in OpenQASM~3 directives and to verify regions larger than ten qubits with equivalence checkers that avoid dense matrices. We also plan to study more quantum compilers and workloads to estimate how often synthesis availability is lost in practice.



\section*{Data Availability}

The replication package, including the data, code, pinned environment, and reproduction instructions, will be made publicly available upon acceptance.

\FloatBarrier

\bibliographystyle{ACM-Reference-Format}
\bibliography{refs}

@article{lintq2024,
  author = {Paltenghi, Matteo and Pradel, Michael},
  title = {Analyzing Quantum Programs with {LintQ}: A Static Analysis Framework for {Qiskit}},
  journal = {Proceedings of the ACM on Software Engineering},
  volume = {1}, 
  number = {FSE},
  pages = {2144--2166},
  year = {2024},
  doi = {10.1145/3660802}
}

@article{openqasm3paper,
  author = {Cross, Andrew and Javadi-Abhari, Ali and Alexander, Thomas and de Beaudrap, Niel and Bishop, Lev S. and Heidel, Steven and Ryan, Colm A. and Sivarajah, Prasahnt and Smolin, John and Gambetta, Jay M. and Johnson, Blake R.},
  title = {{OpenQASM} 3: A Broader and Deeper Quantum Assembly Language},
  journal = {ACM Transactions on Quantum Computing},
  volume = {3},
  number = {3},
  pages = {1--50},
  year = {2022},
  doi = {10.1145/3505636}
}

@misc{openqasmspec,
  author = {{OpenQASM Contributors}},
  title = {{OpenQASM} Live Specification},
  howpublished = {\url{https://openqasm.com}, source repository \url{https://github.com/openqasm/openqasm}},
  year = {2026},
  note = {Language chapters \emph{Directives} and \emph{Gates}; retrieved August 2026}
}

@article{tket2020,
  author = {Sivarajah, Seyon and Dilkes, Silas and Cowtan, Alexander and Simmons, Will and Edgington, Alec and Duncan, Ross},
  title = {t$|$ket$\rangle$: a retargetable compiler for {NISQ} devices},
  journal = {Quantum Science and Technology},
  volume = {6}, number = {1}, pages = {014003}, year = {2020},
  doi = {10.1088/2058-9565/ab8e92}
}

@misc{qiskit2024,
  title = {Quantum computing with {Q}iskit},
  author = {Javadi-Abhari, Ali and Treinish, Matthew and Krsulich, Kevin and Wood, Christopher J. and Lishman, Jake and Gacon, Julien and Martiel, Simon and Nation, Paul D. and Bishop, Lev S. and Cross, Andrew W. and Johnson, Blake R. and Gambetta, Jay M.},
  year = {2024},
  doi = {10.48550/arXiv.2405.08810},
  howpublished = {arXiv:2405.08810}
}

@article{iten2022,
  author = {Iten, Raban and Moyard, Romain and Metger, Tony and Sutter, David and Woerner, Stefan},
  title = {Exact and Practical Pattern Matching for Quantum Circuit Optimization},
  journal = {ACM Transactions on Quantum Computing},
  volume = {3},
  number = {1},
  pages = {1--41},
  year = {2022},
  doi = {10.1145/3498325}
}

@inproceedings{amy2019feynman,
  author = {Amy, Matthew},
  title = {Towards Large-scale Functional Verification of Universal Quantum Circuits},
  booktitle = {Proceedings of the 15th International Conference on Quantum Physics and Logic (QPL)},
  pages = {1--21}, 
  year = {2019},
  doi = {10.4204/EPTCS.287.1}
}

@article{qcpm2024,
  author = {Jiang, Hui and Li, Dian-Kang and Deng, Yu-Xin and Xu, Ming},
  title = {A Pattern Matching Based Framework for Quantum Circuit Rewriting},
  journal = {Journal of Computer Science and Technology},
  volume = {39}, number = {6}, pages = {1312--1327}, year = {2024},
  doi = {10.1007/s11390-024-2726-3}
}

@misc{deqompile2025,
  author = {Xie, Shubing and Sarkar, Aritra and Feld, Sebastian},
  title = {{DeQompile}: Quantum Circuit Decompilation using Genetic Programming for Explainable Quantum Architecture Search},
  howpublished = {arXiv:2504.08310},
  year = {2025},
  doi = {10.48550/arXiv.2504.08310},
}

@article{peham2022zx,
  author = {Peham, Tom and Burgholzer, Lukas and Wille, Robert},
  title = {Equivalence Checking of Quantum Circuits with the {ZX}-Calculus},
  journal = {IEEE Journal on Emerging and Selected Topics in Circuits and Systems},
  volume = {12}, number = {3}, pages = {662--675}, year = {2022},
  doi = {10.1109/JETCAS.2022.3202204}
}

@article{qcec2021,
  author = {Burgholzer, Lukas and Wille, Robert},
  title = {Advanced Equivalence Checking for Quantum Circuits},
  journal = {IEEE Transactions on Computer-Aided Design of Integrated Circuits and Systems},
  volume = {40}, number = {9}, pages = {1810--1824}, year = {2021},
  doi = {10.1109/TCAD.2020.3032630}
}

@misc{llvmtransformmeta,
  author = {{LLVM Project}},
  title = {{LLVM} Documentation: Code Transformation Metadata},
  howpublished = {\url{https://llvm.org/docs/TransformMetadata.html}},
  year = {2026},
  note = {Retrieved August 2026}
}

@misc{llvmnewpm,
  author = {{LLVM Project}},
  title = {{LLVM} Documentation: Using the New Pass Manager},
  howpublished = {\url{https://github.com/llvm/llvm-project}, file \texttt{llvm/docs/NewPassManager.md}},
  year = {2026},
  note = {Section on invalidating analyses; retrieved August 2026}
}

@misc{rosa2025hli,
  author = {Rosa, Evandro C. R. and Marchi, Jerusa and Duzzioni, Eduardo I. and de Santiago, Rafael},
  title = {Optimizing Quantum Compilation via High-Level Quantum Instructions},
  howpublished = {arXiv:2510.24323},
  year = {2025},
  doi = {10.48550/arXiv.2510.24323},
}

@article{rosa2025decomp,
  author = {Rosa, Evandro C. R. and Duzzioni, Eduardo I. and de Santiago, Rafael},
  title = {Optimizing Gate Decomposition for High-Level Quantum Programming},
  journal = {Quantum},
  volume = {9}, pages = {1659}, year = {2025},
  doi = {10.22331/q-2025-03-12-1659}
}

@manual{qirspec,
  author       = {{QIR Alliance}},
  title        = {{QIR Specification}},
  year         = {2021},
  url          = {https://github.com/qir-alliance/qir-spec},
  lastaccessed = {September 10, 2026},
  note         = {Profile specifications under \texttt{specification/profiles}}
}

@misc{jeffspec,
  author = {{Unitary Foundation}},
  title = {jeff: an Interchange Format for Quantum Compilers},
  howpublished = {\url{https://github.com/unitaryfoundation/jeff}},
  year = {2026},
  note = {Specification and schema, v0.3.0}
}

@article{barenco1995,
  author = {Barenco, Adriano and Bennett, Charles H. and Cleve, Richard and DiVincenzo, David P. and Margolus, Norman and Shor, Peter and Sleator, Tycho and Smolin, John A. and Weinfurter, Harald},
  title = {Elementary gates for quantum computation},
  journal = {Physical Review A},
  volume = {52}, number = {5}, pages = {3457--3467}, year = {1995},
  doi = {10.1103/PhysRevA.52.3457}
}

@article{khattar2024,
  author = {Khattar, Tanuj and Gidney, Craig},
  title = {Rise of conditionally clean ancillae for efficient quantum circuit constructions},
  journal = {Quantum},
  volume = {9}, pages = {1752}, year = {2025},
  doi = {10.22331/q-2025-05-21-1752}
}

@article{kms2007,
  author = {Kutin, Samuel A. and Moulton, David P. and Smithline, Lawren M.},
  title = {Computation at a Distance},
  journal = {Chicago Journal of Theoretical Computer Science},
  volume = {2007}, 
  number = {1},
  pages = {1--17},
  year = {2007},
  doi = {10.4086/cjtcs.2007.001}
}

@article{pmh2008,
  author = {Patel, Ketan N. and Markov, Igor L. and Hayes, John P.},
  title = {Optimal synthesis of linear reversible circuits},
  journal = {Quantum Information and Computation},
  volume = {8}, number = {3--4}, pages = {282--294}, year = {2008},
  doi = {10.26421/QIC8.3-4-4}
}

@inproceedings{li2019sabre,
  author    = {Li, Gushu and Ding, Yufei and Xie, Yuan},
  title     = {Tackling the Qubit Mapping Problem for {NISQ}-Era Quantum Devices},
  booktitle = {Proceedings of the 24th International Conference on Architectural Support for Programming Languages and Operating Systems (ASPLOS)},
  pages     = {1001--1014},
  year      = {2019},
  doi       = {10.1145/3297858.3304023}
}

@misc{qiskitqpy,
  author       = {{Qiskit Development Team}},
  title        = {{QPY} Serialization Format},
  howpublished = {Qiskit 2.5.2, module \texttt{qiskit.qpy};
                  \url{https://github.com/Qiskit/qiskit}},
  year         = {2026},
  note         = {Retrieved August 2026}
}

@ArtifactSoftware{qiskithls,
  author       = {{Qiskit Development Team}},
  title        = {{Qiskit} 2.5.2},
  organization = {Qiskit},
  year         = {2026},
  url          = {https://github.com/Qiskit/qiskit/tree/2.5.2/qiskit/transpiler/passes/synthesis},
  lastaccessed = {September 10, 2026},
  note         = {High-level synthesis implementation, including HLS configuration and dispatch and the QFT, MCX, LinearFunction, and PauliEvolution synthesis plugins}
}

@ArtifactSoftware{cirq,
  author       = {{Cirq Developers}},
  title        = {Cirq 1.6.1},
  year         = {2025},
  url          = {https://github.com/quantumlib/Cirq/tree/v1.6.1},
  lastaccessed = {September 10, 2026},
  note         = {Version used in our measurements}
}

@ArtifactSoftware{qiskitqasm3import,
  author       = {Jake Lishman},
  title        = {{qiskit-qasm3-import} 0.6.0},
  organization = {Qiskit},
  year         = {2025},
  url          = {https://github.com/Qiskit/qiskit-qasm3-import/tree/v0.6.0},
  lastaccessed = {September 10, 2026},
  note         = {OpenQASM 3 importer used in the experimental environment}
}

@misc{pytketqiskit,
  author       = {{Quantinuum}},
  title        = {pytket-qiskit: {Qiskit} Extension for pytket},
  howpublished = {\url{https://github.com/Quantinuum/pytket-qiskit}},
  year         = {2026},
  note         = {Version 0.78.0 used in our measurements; retrieved August 2026}
}

@article{chamberland2020,
  author = {Chamberland, Christopher and Zhu, Guanyu and Yoder, Theodore J. and
            Hertzberg, Jared B. and Cross, Andrew W.},
  title = {Topological and Subsystem Codes on Low-Degree Graphs with Flag Qubits},
  journal = {Physical Review X},
  volume = {10}, number = {1}, pages = {011022}, year = {2020},
  doi = {10.1103/PhysRevX.10.011022}
}

@inproceedings{grover1996,
  author = {Grover, Lov K.},
  title = {A Fast Quantum Mechanical Algorithm for Database Search},
  booktitle = {Proceedings of the 28th Annual ACM Symposium on Theory of Computing (STOC)},
  pages = {212--219}, 
  year = {1996},
  doi = {10.1145/237814.237866}
}

@incollection{ghz1989,
  author = {Greenberger, Daniel M. and Horne, Michael A. and Zeilinger, Anton},
  title = {Going Beyond {B}ell's Theorem},
  booktitle = {Bell's Theorem, Quantum Theory and Conceptions of the Universe},
  editor = {Kafatos, Menas},
  publisher = {Springer}, address = {Dordrecht}, pages = {69--72}, year = {1989},
  doi = {10.1007/978-94-017-0849-4_10}
}

@misc{openqasm2,
  author = {Cross, Andrew W. and Bishop, Lev S. and Smolin, John A. and Gambetta, Jay M.},
  title = {Open Quantum Assembly Language},
  howpublished = {arXiv:1707.03429},
  year = {2017},
  doi = {10.48550/arXiv.1707.03429},
}

@inproceedings{paulihedral2022,
  author    = {Li, Gushu and Wu, Anbang and Shi, Yunong and Javadi-Abhari, Ali and Ding, Yufei and Xie, Yuan},
  title     = {Paulihedral: A Generalized Block-Wise Compiler Optimization Framework for Quantum Simulation Kernels},
  booktitle = {Proceedings of the 27th ACM International Conference on Architectural Support for Programming Languages and Operating Systems (ASPLOS)},
  pages     = {554--569},
  year      = {2022},
  doi       = {10.1145/3503222.3507715}
}

@article{assertopt2020,
  author    = {H\"{a}ner, Thomas and Hoefler, Torsten and Troyer, Matthias},
  title     = {Assertion-Based Optimization of Quantum Programs},
  journal   = {Proceedings of the ACM on Programming Languages},
  volume    = {4},
  number    = {OOPSLA},
  pages  = {1--20},
  year      = {2020},
  doi       = {10.1145/3428201}
}

@inproceedings{pnueli1998,
  author = {Pnueli, Amir and Siegel, Michael and Singerman, Eli},
  title = {Translation Validation},
  booktitle = {Tools and Algorithms for the Construction and Analysis of Systems (TACAS)},
  pages = {151--166},
  publisher = {Springer}, year = {1998},
  doi = {10.1007/BFb0054170}
}

@inproceedings{morphq2023,
  author = {Paltenghi, Matteo and Pradel, Michael},
  title = {{MorphQ}: Metamorphic Testing of the {Qiskit} Quantum Computing Platform},
  booktitle = {Proceedings of the 45th International Conference on Software Engineering (ICSE)},
  pages = {2413--2424}, 
  year = {2023},
  doi = {10.1109/ICSE48619.2023.00202}
}

@inproceedings{qdiff2021,
  author = {Wang, Jiyuan and Zhang, Qian and Xu, Guoqing Harry and Kim, Miryung},
  title = {{QDiff}: Differential Testing of Quantum Software Stacks},
  booktitle = {Proceedings of the 36th IEEE/ACM International Conference on
               Automated Software Engineering (ASE)},
  pages = {692--704}, year = {2021},
  doi = {10.1109/ASE51524.2021.9678792}
}

@inproceedings{murali2019noise,
  author    = {Prakash Murali and Jonathan M. Baker and Ali Javadi-Abhari and Frederic T. Chong and Margaret Martonosi},
  title     = {Noise-Adaptive Compiler Mappings for Noisy Intermediate-Scale Quantum Computers},
  booktitle = {Proceedings of the 24th International Conference on Architectural Support for Programming Languages and Operating Systems (ASPLOS)},
  pages     = {1015--1029},
  year      = {2019},
  doi       = {10.1145/3297858.3304075}
}

@online{pytketdocs,
  author       = {{Quantinuum}},
  title        = {{pytket 2.18.1 API documentation}},
  year         = {2026},
  url          = {https://docs.quantinuum.com/tket/api-docs/},
  lastaccessed = {September 10, 2026}
}

@misc{qiskittranspileparams,
  author       = {{IBM Quantum}},
  title        = {Commonly Used Parameters for Transpilation},
  year         = {2026},
  url          = {https://quantum.cloud.ibm.com/docs/en/guides/common-parameters},
  lastaccessed = {September 10, 2026},
  note         = {IBM Quantum Documentation}
}

@misc{qiskit25release,
  author       = {{Qiskit Development Team}},
  title        = {{Qiskit SDK} 2.5 Release Notes},
  year         = {2026},
  url          = {https://quantum.cloud.ibm.com/docs/en/api/qiskit/release-notes/2.5},
  lastaccessed = {September 10, 2026},
  note         = {IBM Quantum Documentation}
}

@misc{goubaultdebrugiere2024faster,
  author = {Goubault de Brugi{\`e}re, Timoth{\'e}e and Martiel, Simon},
  title = {Faster and Shorter Synthesis of Hamiltonian Simulation Circuits},
  year = {2024},
  howpublished = {arXiv:2404.03280},
  doi = {10.48550/arXiv.2404.03280}
}

@article{quetschlich2023mqtbench,
  author  = {Nils Quetschlich and Lukas Burgholzer and Robert Wille},
  title   = {{MQT Bench}: Benchmarking Software and Design Automation Tools for Quantum Computing},
  journal = {Quantum},
  volume  = {7},
  pages   = {1062},
  year    = {2023},
  doi     = {10.22331/q-2023-07-20-1062}
}

@inproceedings{quetschlich2023mqtproblemsolver,
  author    = {Nils Quetschlich and Lukas Burgholzer and Robert Wille},
  title     = {Towards an Automated Framework for Realizing Quantum Computing Solutions},
  booktitle = {Proceedings of the 53rd IEEE International Symposium on Multiple-Valued Logic (ISMVL)},
  pages     = {134--140},
  year      = {2023},
  doi       = {10.1109/ISMVL57333.2023.00035}
}

@inproceedings{vandam2023resourceestimator,
  author    = {Wim van Dam and Mariia Mykhailova and Mathias Soeken},
  title     = {Using Azure Quantum Resource Estimator for Assessing Performance of Fault Tolerant Quantum Computation},
  booktitle = {Proceedings of the SC '23 Workshops of the International Conference on High Performance Computing, Network, Storage, and Analysis (SC-W)},
  pages     = {1414--1419},
  year      = {2023},
  doi       = {10.1145/3624062.3624211}
}

@article{paltenghi2022bugs,
  author    = {Paltenghi, Matteo and Pradel, Michael},
  title     = {Bugs in Quantum Computing Platforms: An Empirical Study},
  journal   = {Proceedings of the ACM on Programming Languages},
  volume    = {6},
  number    = {OOPSLA1},
  articleno = {86},
  pages     = {1--27},
  year      = {2022},
  doi       = {10.1145/3527330}
}

@misc{qamomile,
  key          = {Qamomile},
  author       = {{Jij Inc.}},
  title        = {{Qamomile}: A Quantum Programming {SDK}},
  howpublished = {\url{https://github.com/Jij-Inc/Qamomile}},
  year         = {2026},
  note         = {Project in the studied corpus; pinned commit recorded in the replication package}
}

@misc{qclib,
  author       = {Araujo, Israel F. and Ara{\'u}jo, Ismael C. S. and da Silva, Leon D. and Blank, Carsten and da Silva, Adenilton J.},
  title        = {Quantum computing library (qclib)},
  howpublished = {\url{https://github.com/qclib/qclib}},
  year         = {2025},
  note         = {Version 0.1.15; project in the studied corpus; pinned commit recorded in the replication package}
}

@article{vedral1996,
  author = {Vedral, Vlatko and Barenco, Adriano and Ekert, Artur},
  title = {Quantum networks for elementary arithmetic operations},
  journal = {Physical Review A},
  volume = {54}, number = {1}, pages = {147--153}, year = {1996},
  doi = {10.1103/PhysRevA.54.147}
}

@misc{draper2000,
  author = {Draper, Thomas G.},
  title = {Addition on a Quantum Computer},
  howpublished = {arXiv:quant-ph/0008033},
  year = {2000},
  doi = {10.48550/arXiv.quant-ph/0008033}
}

@misc{cuccaro2004,
  author = {Cuccaro, Steven A. and Draper, Thomas G. and Kutin, Samuel A. and Moulton, David Petrie},
  title = {A new quantum ripple-carry addition circuit},
  howpublished = {arXiv:quant-ph/0410184},
  year = {2004},
  doi = {10.48550/arXiv.quant-ph/0410184}
}

@misc{gidney2017,
  author = {Gidney, Craig},
  title = {Factoring with $n+2$ clean qubits and $n-1$ dirty qubits},
  howpublished = {arXiv:1706.07884},
  year = {2017},
  doi = {10.48550/arXiv.1706.07884}
}

@article{itertestq2026,
  author    = {Paltenghi, Matteo and Pradel, Michael},
  title     = {{IterTestQ}: Assembly-Level, Cross-Platform Testing of Quantum Computing Platforms},
  journal   = {Proceedings of the ACM on Software Engineering},
  volume    = {3},
  number    = {ISSTA},
  articleno = {ISSTA120},
  numpages  = {23},
  year      = {2026},
  doi       = {10.1145/3832211}
}

@misc{sweReplicationPackage2026,
  author = {Boshuai Ye and Peng Liang and Arif Ali Khan},
  title  = {Replication Package for ``When Equivalent Quantum Circuits Lose Synthesis Choices''},
  year   = {2026},
  note   = {Replication package to be made publicly available upon acceptance}
}

\end{document}